# Persistence of correlation-driven surface states in SmB$_6$ under pressure

Soonbeom Seo[1,2,][*], Yongkang Luo[3], S. M. Thomas[1], Z. Fisk[4], O. Erten[5], P. S. Riseborough[6],

F. Ronning[1], J. D. Thompson[1], and P. F. S. Rosa[1]

[1]*Los Alamos National Laboratory, Los Alamos, New Mexico 87545, USA*

[2]*Center for Quantum Materials and Superconductivity (CQMS), Department of Physics, Sungkyunkwan University, Suwon, 16419, Republic of Korea*

[3]*Wuhan National High Magnetic Field Center and School of Physics, Huazhong University of Science and Technology, Wuhan 430074, China*

[4]*Department of Physics and Astronomy, University of California, Irvine 92697, USA*

[5]*Department of Physics, Arizona State University, Tempe, AZ 85281, USA*

[6]*Department of Physics, Temple University, Philadelphia, PA 19122, USA*

**Abstract**

The proposed topological Kondo insulator SmB$_6$ hosts a bulk Kondo hybridization gap that stems from strong electronic correlations and a metallic surface state whose effective mass remains disputed. Thermopower and scanning tunnelling spectroscopy measurements argue for heavy surface states that also stem from strong correlations, whereas quantum oscillation and angle-resolved photoemission measurements reveal light effective masses that would be consistent with a Kondo breakdown scenario at the surface. Here we investigate the evolution of the surface state via electrical and thermoelectric transport measurements under hydrostatic pressure, a clean symmetry-preserving tuning parameter that suppresses the Kondo gap and increases the valence of Sm from ~2.6+ towards a 3+ magnetic metallic state. Electrical resistivity measurements reveal that the surface carrier density increases with increasing pressure, whereas thermopower measurements show an unchanged Fermi energy under pressure. As a result, the effective mass of the surface state charge carriers linearly increases with pressure as the Sm valence approaches 3+. Our results are consistent with the presence of correlation-driven surface states in $SmB_6$ and suggest that the surface Kondo effect persists under pressure to 2 GPa.

**INTRODUCTION**

$SmB_6$ belongs to a family of materials known as Kondo insulators in which a small gap in the electronic spectrum at the Fermi energy ($E_F$) arises from hybridization of weakly dispersing $f$-states and more strongly dispersing conduction band states provided by $d$-electrons [1,2]. Strong electronic correlations control physical properties, including the quantum admixture of nearly degenerate $f^n$ and $f^{n+1}$ valence configurations that leads to an intermediate valence state [3,4]. In the case of $SmB_6$, the nominal valence of Sm is about 2.6 at room temperature and atmospheric pressure [5–7]. If parity of the occupied bands changes an odd number of times at high-symmetry points in the Brillouin zone, topologically protected surface states are predicted to emerge and give rise to a topological Kondo insulator (TKI) phase of matter [8,9].

Experimental observations using non-local [10] and geometry-dependent [11–13] transport measurements on $SmB_6$ have confirmed the existence of the metallic surface state in $SmB_6$. These surface states are a direct consequence of the presence of a correlation-driven gap in the bulk, and their robustness against non-magnetic disorder and heavy-ion irradiation has been probed experimentally [14–16]. Further, angle-resolved photoemission spectroscopy (ARPES) [17–19] and scanning tunneling microscopy (STM) [20] measurements on $SmB_6$ support the presence of a topologically ordered metallic surface state, consistent with theoretical expectations [21,22]. Even though several ARPES and STM studies argue against the topological nature of $SmB_6$ [23–25], numerous recent experimental and theoretical studies have strongly suggested the topologically non-trivial nature of SmB$_6$ [26–28]. Theoretically, the effective mass (m*) of surface charge carriers depends on the hybridization strength at the surface [29–31] and is inversely proportional to the carriers' Fermi velocity ($v_F$). Experimentally, conflicting conclusions have been reported. Analysis of quantum oscillation [32] and ARPES [18,19,33] measurements is consistent with a small effective mass (m*/m$_e$ = 0.1 ~ 1) of the surface state carriers, which is at odds with a large effective mass (m* of order a few 100m$_e$) deduced from thermoelectric power, Nernst [34], and scanning tunneling spectroscopy [27,35,36]. Even the assignment of quantum oscillations to surface states in flux-grown SmB$_6$ remains a matter of debate [37–39].

A related open question is the evolution of the surface state and m* as a function of applied pressure. Electrical transport and x-ray spectroscopy studies of SmB$_6$ reveal that the bulk insulating state transits to a bulk metal state under pressure on the order of 5-10 GPa and is accompanied by a change of the valence of Sm ions towards a trivalent configuration [40,41]. Though the precise magnitude of the critical pressure for bulk metallization depends on the hydrostaticity of the pressure medium [42,43], applying pressure generically is expected to decrease the strength of renormalized, effective hybridization [44,45] in SmB$_6$ as its ground state evolves toward magnetic order with an integer valence. Microscopic probes, such as ARPES and scanning tunneling spectroscopy, are not available under high pressure conditions, but thermopower experiments, which are sensitive to m* and the bulk gap, are accessible.

Here, we investigate the evolution of the surface state of SmB$_6$ under hydrostatic pressure. Thermoelectric power reveals that the Fermi energy is essentially unchanged by pressures to 2 GPa. Electrical resistivity under pressure provides surface and bulk contributions using a two-channel model. Combining thermoelectric and electrical transport results, we find that the effective mass of the surface state charge carriers in SmB$_6$ linearly increases with increasing pressure. Our results, an initial step in determining the effect of pressure on the surface states of SmB$_6$, place constraints on theoretical predictions and point to the need for measurements to higher pressures to track the evolution of the surface state as the bulk becomes metallic.

**METHODS**

Single crystals of SmB$_6$ were synthesized by an aluminum-flux method described in Ref. [16]. The orientation of the polished sample was verified by Laue diffraction at room temperature. Pressure was generated in a hybrid Be–Cu/NiCrAl clamp-type pressure cell with Daphne oil 7373 as a pressure-transmitting medium to ensure hydrostatic condition for pressure up to 2 GPa. Pressure in the cell was determined from the pressure-dependent superconducting transition temperature of a Pb manometer using the pressure scale of Eiling and Schilling [46]. A standard four-probe technique was employed to measure the electrical resistivity, using a Lakeshore Model 372 AC Resistance Bridge. A steady-state technique was used to measure the thermopower, where a longitudinal thermal gradient on the sample is induced by a heater and a heat sink attached to the sample. A pair of well-calibrated differential chromel-Au/Fe thermocouples and a pair of platinum wires were used to measure the temperature gradient and thermal voltage, respectively. The thermopower was measured after the sample reached the steady state at fixed temperatures. Both electrical resistivity and thermopower measurements were made on the same crystal. Two different cryostats were used to control temperature: a $^4$He cryostat for temperature measurements from 300 to 1.8 K and a $^3$He cryostat for temperatures from 50 down to 0.3 K.

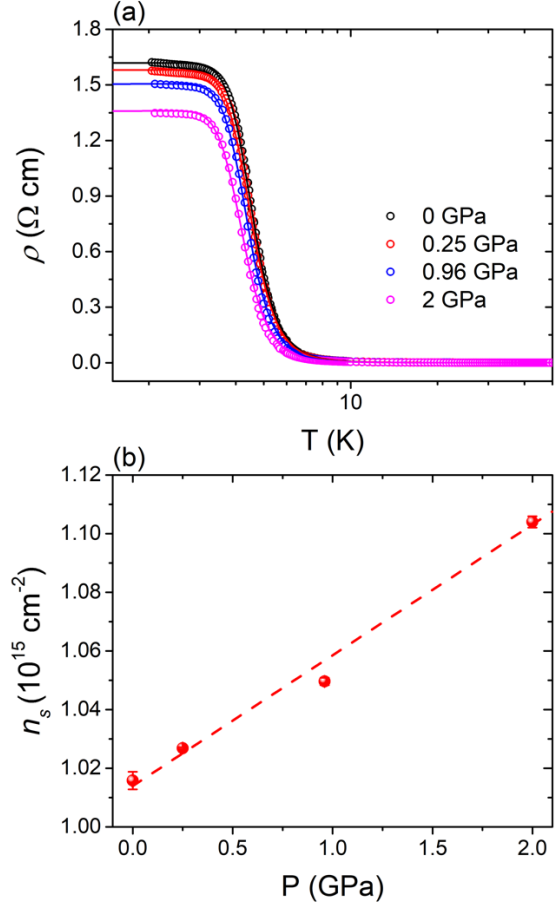

Figure 1 (a) Temperature dependence of the electrical resistivity on the (001) plane of $SmB_6$ under applied pressure. Solid lines are fits to a two-channel model described in the text. (b) Pressure dependence of the carrier density $n_s$ of the surface state in $SmB_6$ from the two-channel model fits.

**RESULTS**

Figure 1(a) shows the temperature dependence of the electrical resistivity on the (001) plane of $SmB_6$ under applied pressure to 2 GPa. The resistivity increases with decreasing temperature for all pressures, in agreement with an insulating response arising from opening a hybridization gap in the bulk. Below $T^* \sim 4$ K, electrical transport is dominated by the surface state contribution, and the resistivity saturates [11,47]. With increasing pressure, both $T^*$ and the low-temperature residual resistivity decrease, in agreement with the previous results [40,43]. To investigate the evolution of surface and bulk contributions under pressure, we fit the resistivity data up to 50 K to a two-channel model, wherein the zero-field diagonal conductivity can be described in terms of surface and bulk contributions as:

$$\sigma_{xx} = \frac{2|e|n_s}{t}\mu_s(T) + n_b(T)|e|\mu_b. \qquad (1)$$

Here $t$ is the thickness of the sample, $e$ is the charge of an electron, $n$ and $\mu$ are carrier density and mobility, respectively, and the subscripts $s$ and $b$ denote the surface and bulk contributions, respectively. We follow previous reports at ambient pressure [34] and assume that $\frac{1}{\mu_s(T)} = \frac{1}{\mu_{s0}}(1 + cT^\gamma)$ according to Matthiessen's rule, and $n_b(T) = n_{b0}\exp(-\Delta_b/T)$. Here $n_s, \mu_{s0}, c, \gamma, n_{b0}, \Delta_b$, and $\mu_b$ are free parameters in a fit to the temperature dependent resistivity. Our results, shown as solid lines in Fig. 1(a), reveal that electrical resistivity under pressure can be well reproduced by the two-channel model. In particular, our fits provide two distinct contributions for all pressures: the surface contribution dominates at low temperature, whereas the bulk contribution is dominant at high temperature. All parameters at ambient pressure are comparable to those of Ref. [34], which argued for heavy surface states in SmB$_6$, even though the surface planes, as well as surface conditions, are different in the two measurements. For the surface-dominated contributions, we find that parameters $c$ and $\gamma$ are small (<< 1) for all pressures, indicating a temperature-independent surface state. Figure 1(b) shows the pressure dependence of the carrier density $n_s$ of the surface state assuming pressure-independent scattering. At ambient pressure $n_s \approx 1.02 \times 10^{15}$ $cm^{-2}$ and it linearly increases by about 9% from ambient pressure to 2 GPa. We highlight, however, that the

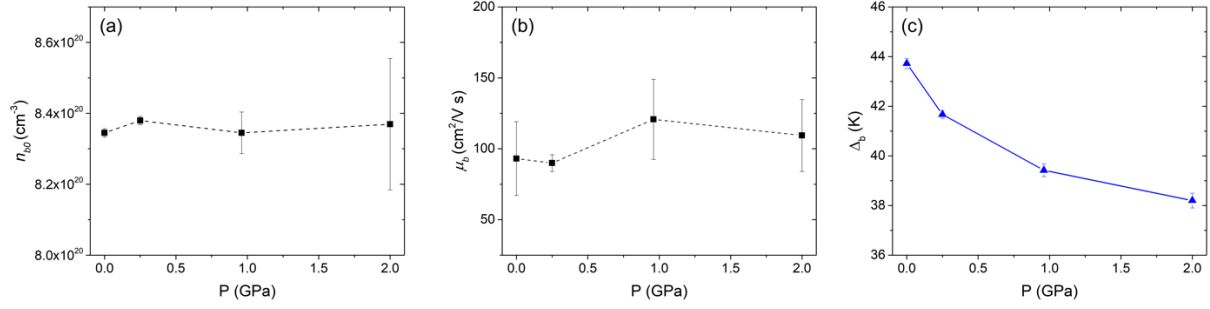

Figure 2 (a) and (b) Pressure dependence of the carrier density $n_{b0}$ and mobility $\mu_b$ of the bulk contribution from the two-channel model fits. Error bars are representative. (c) Pressure dependence of the bulk energy gap parameter $\Delta_b$ from a two-channel model fit.

diagonal conductivity depends on the product between the carrier density, $n_s$, and the mobility, $\mu_{s0}$, and it is difficult to disentangle these two parameters based solely on fits to the electrical resistivity. To further confirm this result, we therefore use the same two-channel model to fit simultaneously $\sigma_{xx}$ and the off-diagonal (Hall) conductivity measured in a previous report as a function of pressure [41]. Our fit results show a similar rate of increase of $n_s$ under pressure (details can be found in the Appendix). This increase can be seen directly from the Hall coefficient data at low temperature, which reveal that the inverse Hall coefficient $1/R_H$ ($\propto n_s$) increases with increasing pressure [40,41]. We note that the magnitude of extracted $n_s$ at ambient pressure depends on the measurements and surface conditions [48–50]. In particular, an upper carrier density limit of 5.85 x $10^{14}$ cm$^{-2}$ is expected from the size of the surface-Brillouin zone of SmB$_6$ within a single-band model. Our carrier density values are 1.7 times larger than this upper limit, and additional conduction channels along unpolished side surfaces are a plausible cause of this discrepancy [48]. Nevertheless, we emphasize that, though the magnitude of the carrier densities in our fits may not be quantitatively accurate and the simplifying assumption of independent surface and bulk conduction channels may not be rigorously valid, the pressure dependences of $n_s$ are consistent across the different samples, measurements, and pressure cells.

From the activated bulk contribution at high temperatures, we cannot obtain an unambiguous pressure dependence of carrier density $n_{b0}$ and mobility $\mu_b$ as shown in Fig. 2(a) and (b). Nonetheless, the bulk gap $\Delta_b$ shows the expected pressure dependence [42,43], *i.e.*, $\Delta_b$ monotonically decreases with increasing pressure as shown in Fig. 2(c).

Next, we investigate the evolution of thermoelectric transport of SmB$_6$ under pressure, which is sensitive to the Fermi surface morphology [34]. Figure 3(a) shows the temperature dependence of thermopower $S$ under pressure, which is measured on the same surface as the electrical resistivity described above. For all pressures, $S(T)$ is negative and displays a maximum around 5 K before its magnitude is quickly suppressed. Above 5 K, the magnitude of thermopower decreases with increasing temperature, roughly following a 1/T dependence, which is characteristic of thermal activation of carriers across a bulk gap in an intrinsic semiconductor. Below $T^*$, $S(T)$ displays a shallow T-linear slope as it approaches zero for all pressures. This response is characteristic of a metal with heavy charge carriers [51]. In agreement with resistivity results, $T^*$ decreases with increasing pressure. Remarkably, the slope of $S(T)$ remains unchanged within experimental accuracy for all pressures ($S/T \approx$ -6.0(2) $\mu$V/K$^2$) as indicated by the dashed guideline in the inset of Fig. 3(a). Because $S/T$ is inversely proportional to the Fermi energy [34,51,52], our results reveal that within experimental uncertainty $\varepsilon_F$ remains constant in SmB$_6$ under pressures to 2 GPa.

In order to probe the evolution of the surface state under pressure, we combine electrical and thermoelectric transport results. Using the pressure dependence of $n_s$ from resistivity data and of $\varepsilon_F$ from thermopower data, an effective mass of SmB$_6$ for the surface state under pressure is

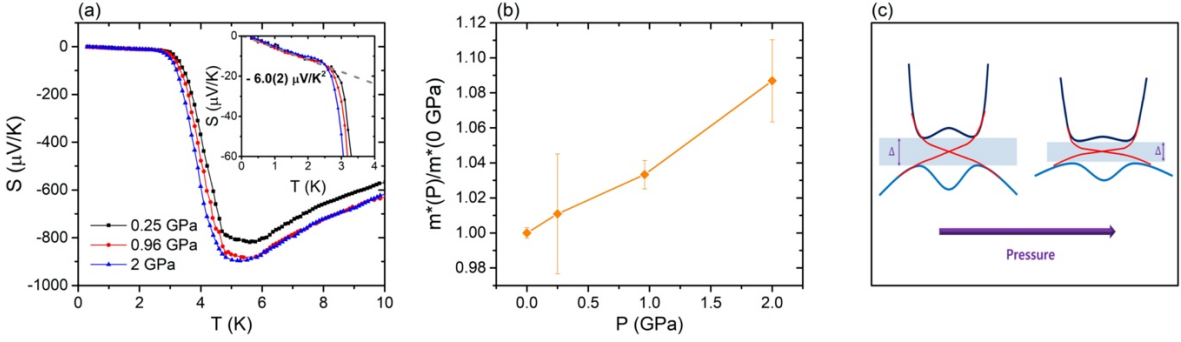

Figure 3 (a) Temperature dependence of thermopower, $S$, for SmB$_6$ at pressures to 2 GPa. The inset is a magnified view below crossover temperature $T^*$ where the surface contribution dominates. The dashed line indicates the slope of thermopower at low temperature with $S/T \approx -6.0(2)$ $\mu$V/K$^2$. (b) Pressure dependence of the effective mass of SmB$_6$ surface state quasiparticles, normalized by its value at ambient pressure. (c) Schematic sketch of the band structures of SmB$_6$ at ambient pressure (left) and high pressure (right). Valence and conduction bands remain qualitatively unchanged, but the energy gap $\Delta$ is reduced with pressure. Red lines indicate band dispersion for surface states within the bulk insulating gap. For purposes of illustration, the relative flattening of bulk and surface states under pressure is exaggerated.

calculated from the relation for a free-electron gas in two dimensions $m^* = \frac{\hbar^2 k_F^2}{2\varepsilon_F} = \frac{2\pi\hbar^2 n_s}{\varepsilon_F}$. Figure 3(b) shows the pressure dependence of the effective mass, normalized by its value at ambient pressure. The linear increase of m* with pressure reaches ~10% at 2 GPa, which indicates that the surface states in SmB$_6$ become heavier with increasing pressure.

**DISCUSSION**

Tuning correlation-driven topological matter by symmetry-preserving parameters may provide new physical insights into topologically-protected surface states. In SmB$_6$, the effective hybridization, i.e., bare hybridization renormalized by the on-site *f*-electron repulsion [44], is expected to be reduced under pressure as the intermediate valence state of Sm is driven toward a trivalent state that gives rise to an antiferromagnetic metallic ground state [42,43]. In a scenario where the surface experiences a Kondo breakdown, a strong reduction in effective hybridization at the surface would be expected to give rise to surface states with a large Fermi surface and light quasiparticles [30,31]. Our electrical and thermoelectric transport results, however, do not show evidence for a surface Kondo breakdown either at ambient pressure or under pressures up to 2 GPa as the heavy effective mass of the surface state charge carriers in SmB$_6$ increases linearly with increasing pressure. Higher pressures may be needed to achieve a surface Kondo breakdown regime. In general, hybridization should be weaker at the surface than in the bulk due to a reduced surface-coordination number [31], and consequently, higher pressures may drive the surface to a Kondo breakdown limit, liberating light quasiparticles, before completely closing the bulk correlation-driven gap.

The surface state evolution at low pressure invites further consideration of the underpinning mechanism driving the pressure dependence of the effective mass. An Anderson-lattice framework may provide insight into the pressure dependence of SmB$_6$ due to its intermediate valence. One possibility is that the effective mass of the surface states simply follows the increase in renormalization of bulk bands. Within the mean-field approximation to the periodic Anderson model (PAM), the scaling expression $m^*/m = (\Delta/T_K)^2$ yields a linear relationship between the effective mass and the square of the ratio of a direct hybridization gap $\Delta$ to the Kondo temperature, $T_K$, [44] a relation verified experimentally in several heavy-fermion compounds [53,54]. The direct gap Δ, determined by the renormalized hybridization strength $\tilde{V}$, is proportional to ~ $(T_K W)^{1/2}$, and to $(\Delta_i W/2)^{1/2}$ where $W$ is the bare conduction bandwidth and $\Delta_i$ is the smaller indirect hybridization gap [44] to which transport measurements are sensitive. As a result, the effective mass of renormalized bands in the bulk increases as a function of pressure due to the reduction in effective hybridization and consequently $T_K$. We note, however, that the PAM relationships above are generically valid for a spin-1/2 electron and have been experimentally tested in Ce and Yb compounds [53]. In SmB$_6$, larger degeneracies are present, and details of the *f*-band dispersion may be important. As reported in Ref. [55], however, M. Haverkort pointed out that for determining the symmetry and number of the single particle 4*f* excitations nearest the chemical potential, the entirely filled spin-orbit split *j* = 5/2 orbitals of divalent (4$f^6$) Sm, with lowest energy atomic multiplet $^7F_{J=0}$, can act as a filled shell, analogous in this respect to the 4$f^0$ and 4$f^{14}$ manifolds for Ce and Yb, respectively. Thus, one could hypothesize that the PAM relations may also be applicable in SmB$_6$.

Our findings that the hybridization gap decreases and the Fermi energy shifts little, if at all, with applied pressure lead to the schematic representation sketched in Fig. 3(c) wherein the surface state linear energy dispersion is confined within the insulating bulk energy gap Δ. This schematic representation is supported by previous ARPES and STM measurements [18,19,33,56] that report a linear energy dispersion of surface bands in the bulk band gap of SmB$_6$ as predicted theoretically [22,57,58]. In particular, a recent scanning tunneling microscopy experiment imaged the formation of linearly dispersing surface states with heavy Dirac fermions, namely m*=(410±20)$m_e$ determined by the Fermi velocity at the Dirac point [27]. As illustrated in Fig. 3(c), a shallower dispersion of surface states within the narrower bulk insulating gap under pressure (red lines) naturally yields a reduction in the Fermi velocity ($v_F$) of the surface states. The resulting effective mass, $m^* = \hbar k_F/v_F$, is therefore enhanced. Even though further investigations are needed to verify the evolution of the energy dispersion of the surface states under pressure, this picture is generically applicable for understanding the origin of a pressure-induced enhancement of $m^*$ on the surface, and it is consistent with expectations of the periodic Anderson model when the effective mass of the surface states tracks the evolution of $m^*$ of renormalized bands in the bulk.

**CONCLUSION**

In summary, combining pressure-dependent thermoelectric and electrical transport measurements, we find that the effective mass of the surface-state quasiparticles linearly increases with increasing pressure. This is inconsistent with a complete surface Kondo-breakdown scenario, which is expected to lead to surface states possessing a large Fermi surface of light quasiparticles, but may be understood within a periodic Anderson model framework. Our results, an initial step in determining the effect of pressure on the surface states of $SmB_6$, place constraints on theoretical predictions and point the need for measurements to higher pressures to track the evolution of the surface state as the bulk is driven close to a metal-insulator, non-magnetic/magnetic boundary.

**Acknowledgments**

We acknowledge constructive discussions with P Coleman and J Allen. We also thank L Sun for kindly sharing Hall data from Ref. [41]. Work at Los Alamos was performed under the auspices of the U.S. Department of Energy, Office of Basic Energy Sciences, Division of Materials Science and Engineering. SS acknowledges support from the Laboratory Directed Research and Development program, and by the Basic Science Research Program through the National Research Foundation of Korea (NRF) funded by the Ministry of Education (NRF-2020R1A6A3A01099912).

**APPENDIX: PRESSURE DEPENDENCE OF THE SURFACE CARRIER DENSITY**

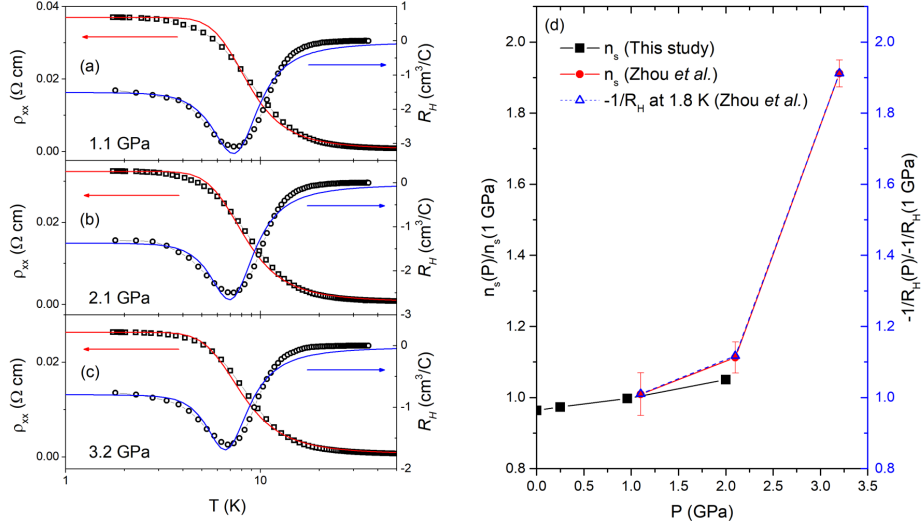

Figure 4. (a-c) Temperature dependence of resistivity $\rho_{xx}$ (squares) and Hall coefficient $R_H$ (circles) plotted on the left and right ordinates, respectively, for pressures of 1.1 GPa (a), 2.1 GPa (b), and 3.2 GPa (c). The solid lines are fits to a two-channel model. Data are from Zhou et al. in Ref. [41]. (d) Pressure dependence of the carrier density $n_s$ of the surface state and the inverse Hall coefficient $-1/R_H$ plotted on the left and right ordinates, respectively. Data are normalized at P = 1 GPa. The solid squares indicate $n_s$ in this study. Error bars are smaller than the size of the data points. The solid circles indicate $n_s$ from the two-channel model fit, and the open triangles indicate $-1/R_H$ at 1.8 K taken from Fig. 4(a-c).

In this Appendix, we fit the conductivity from Ref. [41] using the two-channel model. Figures 4(a-c) show the temperature dependence of the electrical resistivity $\rho_{xx}$ and Hall coefficient $R_H$ of SmB$_6$ for pressures at 1.1, 2.1, and 3.2 GPa, plotted on the left and right ordinates, respectively. The data were obtained from Ref. [41]. We fit these data to a two-channel model, wherein the diagonal conductivity $\sigma_{xx}$ is described by equation (1) in the text and off-diagonal conductivity $\sigma_{xy}$ is:

$$\sigma_{xy} = \frac{2en_s}{t}\frac{\mu_s^2(T)B}{1+\mu_s^2(T)B^2} + n_b(T)e\frac{\mu_b^2 B}{1+\mu_b^2 B^2}$$

where $t$ is the thickness of the sample, $e$ is the charge of an electron, $n$ and $\mu$ are carrier density and mobility, respectively, and the subscripts $s$ and $b$ denote the surface and bulk contributions, respectively. We obtained common parameters which fit both diagonal ($\rho_{xx}$) and off-diagonal ($\rho_{yx} = R_H B$, $B = 1$ T) resistivity data as indicated by solid lines in Fig. 4(a-c). Figure 4(d) shows the pressure dependence of the carrier density $n_s$ of the surface state and the inverse Hall coefficient - $1/R_H$ at 1.8 K plotted on the left and right ordinates, respectively. Data are normalized at P = 1 GPa. The solid squares and circles indicate $n_s$ from the two-channel model fit of our resistivity data in the text and in the resistivity and Hall coefficient data in Ref. [41] (Fig. 4(a-c)), respectively. The pressure dependence of both normalized $n_s$s are similar to each other up to 2 GPa. Above 2 GPa, $n_s$ from Zhou et al. is significantly enhanced. To further confirm this result, we also plot the pressure dependence of normalized inverse Hall coefficient - $1/R_H$ at 1.8 K as shown in Fig. 4(d). Since inverse Hall coefficient at low temperature is dominated by the surface carrier density of $SmB_6$, the observation that the pressure dependence of normalized - $1/R_H$ is consistent with that of $n_s$ from the two-channel model supports the conclusion that the carrier density of the surface state increases with pressure. A previous report showed that $-1/R_H$ at 1.7 K in $SmB_6$ rapidly increases at higher pressure, particularly near a critical pressure where the bulk insulating state transits to a bulk metal state [40].